\documentclass[twocolumn]{aastex62}

\usepackage{graphicx}	% Including figure files
\usepackage{amsmath}	% Advanced maths commands
\usepackage{amssymb}	% Extra maths symbols

\received{November 11, 2019}
\accepted{---}
\submitjournal{ApJL}

\shorttitle{[\ion{O}{1}]63-um detection at $z\geq6$}
\shortauthors{Rybak et al.}

\begin{document}

\title{First detection of the [\ion{O}{1}] 63-$\mu$m emission from a redshift 6 dusty galaxy}

\correspondingauthor{Matus Rybak}
\author[0000-0002-1383-0746]{Matus Rybak}
\email{mrybak@strw.leidenuniv.nl}
\affiliation{Leiden Observatory, Leiden University, Niels Bohrweg 2, 2333 CA Leiden, the Netherlands}

\author{J. A. Zavala}
\affiliation{Department of Astronomy, The University of Texas at Austin, 2515 Speedway Blvd Stop C1400, Austin, TX 78712, USA}

\author{J. A. Hodge}
\affiliation{Leiden Observatory, Leiden University, Niels Bohrweg 2, 2333 CA Leiden, the Netherlands}

\author{C. M. Casey}
\affiliation{Department of Astronomy, The University of Texas at Austin, 2515 Speedway Blvd Stop C1400, Austin, TX 78712, USA}

\author{P. van der Werf}
\affiliation{Leiden Observatory, Leiden University, Niels Bohrweg 2, 2333 CA Leiden, the Netherlands}

\begin{abstract}
 We report a ground-based detection of the [\ion{O}{1}] 63-$\mu$m line in a $z=6.027$ gravitationally lensed dusty star-forming galaxy (DSFG) G09.83808 using the APEX SEPIA~660 receiver, the first unambiguous detection of the [\ion{O}{1}]$_{63}$ line beyond redshift 3, and the first obtained from the ground.
 The [\ion{O}{1}]$_{63}$ line is robustly detected at 22$\pm$5~Jy km s$^{-1}$, corresponding to an intrinsic (de-lensed) luminosity of $(5.4\pm1.3)\times10^{9}$~L$_\odot$. With the [\ion{O}{1}]$_{63}$/[\ion{C}{2}] luminosity ratio of 4, the [\ion{O}{1}]$_{63}$ line is the main coolant of the neutral gas in this galaxy, in agreement with model predictions. The high [\ion{O}{1}]$_{63}$ luminosity compensates for the pronounced [\ion{C}{2}] deficit ([\ion{C}{2}]/FIR$\simeq4\times10^{-4}$). Using photon-dominated region models, we derive a source-averaged gas density $n=10^{4.0}$~cm$^{-3}$, and far-UV field strength $G=10^4~G_0$, comparable to the $z=2-4$ DSFG population. If G09.83808 represents a typical high-redshift DSFG, the [\ion{O}{1}]$_{63}$ line from $z=6$ non-lensed DSFGs should be routinely detectable in ALMA Band~9 observations with $\sim$15 min on-source, opening a new window to study the properties of the earliest DSFGs.

\end{abstract}


\section{Introduction} \label{sec:intro}

Although thousands of the sub-millimeter bright, dusty star-forming galaxies (DSFGs) have been discovered at $z=2-5$ (e.g.,\citealt*{Casey2014}), the number of known DSFGs drops precipitously at $z\geq5$: only a handful of $z\geq6$ DSFGs have been discovered to-date \citep{Riechers2013, Decarli2017, Strandet2017,Zavala2018}. These dust-laden sources provide evidence for intense star-formation and interstellar medium (ISM) enrichment within the first Gyr of cosmic history, and extremely efficient baryon conversion. Characterizing the conditions of their star-forming ISM - particularly the gas density of the star-forming clouds and the FUV radiation field illuminating them - is a key to understanding these extreme sources.

Far-IR fine-structure lines of [\ion{C}{2}], [\ion{O}{1}] and  [\ion{C}{1}] and the CO rotational lines are the key diagnostics of the neutral and molecular gas in the star-forming clouds. By comparing the observed line and continuum fluxes to photochemical models, the ISM properties such as the gas density ($n$) and the strength of the incident FUV radiation ($G$) can be inferred. Indeed, CO and [\ion{C}{2}] lines have been instrumental in studying the ISM of $z=2-5$ DSFGs (e.g., \citealt{Stacey2010, Gullberg2015, Wardlow2017, Zhang2018b, Rybak2019}) down to sub-kpc scales \citep{Lamarche2018, Yang2019, Rybak2019b}; these have revealed a dense ISM (n=$10^3-10^5$ cm$^{-3}$) exposed to strong FUV fields ($G=10^2-10^5~G_0$)\footnote{The far-UV field strength is given in Habing field units, $1\,G_0 = 1.6\times10^{-3}$ erg s$^{-1}$ cm$^{-2}$, a typical value for the Galactic interstellar FUV field.}.

At $z\geq5$, our toolkit for studying the neutral star-forming ISM becomes much more limited. While the ALMA has enabled routine studies of the [\ion{C}{2}] 158-$\mu$m (e.g., \citealt{Decarli2017, Smit2018}) and [\ion{O}{3}] 88-$\mu$m emission (e.g., \citealt{Inoue2016, Carniani2017, Hashimoto2019, Harikane2019}), the former arises from both the ionized and neutral ISM, while the latter is associated with \ion{H}{2} regions. The low-$J$ CO lines become extremely difficult to detect due to both their intrinsic faintness and the elevated cosmic microwave background (CMB) temperature (e.g., \citealt{daCunha2013}). Although mid-$J$ CO lines remain detectable at $z\geq5$, their interpretation is sensitive to the details of radiative transfer assumptions (e.g., optical depth and turbulence, \citealt{Popping2019}) and the CMB background \citep{daCunha2013}. 

However, in a dense, warm ISM - such as that in DSFGs or present-day (ultra)luminous infrared galaxies (ULIRGs) - the [\ion{O}{1}] 63-$\mu$m line ([\ion{O}{1}]$_{63}$) overtakes [\ion{C}{2}] as the main gas cooling channel \citep{Kaufman1999, Kaufman2006, Narayanan2017}. With a critical density $n_\mathrm{crit}\simeq5\times10^5$~cm$^{-3}$, [\ion{O}{1}]$_{63}$ traces much denser ISM than the [\ion{C}{2}] emission ($n_\mathrm{crit}=3\times10^3$~cm$^{-3}$ for collisions with hydrogen in PDRs). Indeed, cosmological hydrodynamical simulations (e.g., \citealt{Olsen2017, Katz2019}) predict [\ion{O}{1}]$_{63}$ to be the most luminous FIR line in star-forming galaxies at the highest redshifts. Unlike CO emission, the [\ion{O}{1}]$_{63}$ line is not strongly affected by the CMB background and local excitation conditions; and unlike [\ion{C}{2}], it is directly associated with the neutral ISM. 

Ground-based studies of the [\ion{O}{1}]$_{63}$ emission at $z\geq1$ have been limited by the atmospheric absorption at sub-mm wavelengths. Above the atmosphere, the [\ion{O}{1}]$_{63}$ emission from $z\sim0$ (ultra) luminous infrared galaxies (ULIRGs) has been extensively studied with ISO \citep{Brauher2008} and \textit{Herschel} \citep{GraciaCarpio2011, Herrera2018, Diaz2017}. Unfortunately, at $z\geq1$, the limited collecting area and on-source time resulted in only $\sim15$ [\ion{O}{1}]$_{63}$ detections \citep{Ivison2010, Sturm2010, Brisbin2015, Coppin2012, Wardlow2017, Zhang2018b}, mainly in gravitationally lensed galaxies, and only out to $z\simeq3$ \citep{Zhang2018b}. However, at $z\geq5.5$, [\ion{O}{1}]$_{63}$ is redshifted into ALMA Band~10, and at $z\geq6.0$, into ALMA/APEX Band~9, making it observable from the ground. In this Letter, we report the first ground-based detection of the [\ion{O}{1}]$_{63}$ line from a $z=6.027$ strongly lensed DSFG, achieved using APEX SEPIA~660 spectroscopy.

\section{Observations}

We targeted G09.83808 (J2000 09:00:45.8 +00:41:23), a $z=6.027$ strongly gravitationally lensed DSFG\footnote{Adopting a flat $\Lambda$CDM cosmology from \citet{Planck2015}, $z=6.027$ corresponds to a luminosity distance of 59350~Mpc, and the age of Universe of 0.94~Gyr \citep{Wright2006}.}, discovered in the \textit{Herschel} H-ATLAS survey. \citet{Zavala2018} obtained a robust spectroscopic confirmation from [\ion{C}{2}] (Sub Millimeter Array, SMA) and CO (5--4)/(6--5) and H$_2$O lines (Large Millimeter Telescope, LMT). Using high-resolution ALMA Band~7 imaging, \citet{Zavala2018} confirmed that G09.83808 is strongly gravitationally lensed, with a FIR magnification $\mu_\mathrm{FIR}\simeq9$. Based on the FIR and mm-wave spectroscopy, G09.83808 has a source-plane FIR luminosity $L_\mathrm{FIR}=(3.8\pm0.5)\times10^{12}~L_\odot$ (8--1000~$\mu$m), corresponding to a star-formation rate SFR of $\sim650$~M$_\odot$ yr$^{-1}$ (assuming the Salpeter initial mass function, \citealt*{Kennicutt1998}). Due to its strongly lensed nature and fortuitous redshift, G09.83808 is ideally suited for [\ion{O}{1}]$_{63}$ observations.

The observations were carried out using the Atacama Pathfinder EXperiment (APEX) 12-m telescope, and the Swedish ESO PI (SEPIA) Band~9 receiver \citep{Belitsky2018, Hesper2017, Hesper2018}, as a part of the NOVA Guaranteed Time Observations (Proposal 0104.B-0551, PI: Rybak).

The observations were carried out in two blocks: 2019 October 28 (5.6~h total time, 97~min on-source, source elevation 41--66~deg) and 2019 November 6 (2.6~h total time, 36~min on-source, source elevation 39--70~deg). 

The observations were conducted in an on/off mode, with the secondary wobbler frequency of 1.5 Hz. For the Oct 28 observations, the initial pointing and calibration was done on R~Dor; for the Nov 6 observations, using o-Ceti. The band-pass calibration and intermediate calibration and pointing checks were performed using IRC+10216 on both dates. Two scans on Oct 28 were aborted due to tracking errors.

The observing conditions were excellent, with the precipitable water vapour of 0.45--0.55~mm (Oct 28) and 0.30--0.35~mm (Nov 6), corresponding to an atmospheric transmission of 0.6--0.8 at 675.2~GHz. The total observing time was 8.2~h, with 133~min on-source time.

The frequency setup consisted of two sidebands, each consisting of two spectrometers with 4096 channels 0.9765 MHz (0.43~km s$^{-1}$) wide, giving a total bandwidth of 8~GHz per sideband. For both the October 28 and November 6 observations, we used two separate tunings with the line-containing spectrometer centered at 673.920~GHz (44.3~min on-source) and 674.920~GHz GHz (53.1~min on-source), respectively. 

At the observed line frequency of 675.220~GHz, the APEX primary beam full-width at half maximum (FWHM) is 9.2~arcsec, compared to the G09.83808 image separation of $\sim2$~arcsec. Although DSFGs show high multiplicity (e.g., \citealt{Hodge2013, Decarli2017}), the high-resolution SMA ($\sim2$~arcsec) and ALMA imaging ($\sim1.2$~arcsec) did not detect and FIR- or [\ion{C}{2}] bright companion source to G09.83808. The observed [\ion{O}{1}]$_{63}$ emission can be thus unambiguously assigned to G09.83808.

The data was reduced using the {\sc Gildas} CLASS package\footnote{\texttt{http://www.iram.fr/IRAMFR/GILDAS/}}. Each tuning was processed separately, before combining the data. The two linear polarizations were combined into the Stokes $I$. After windowing the channels containing the line or the atmospheric lines, we subtract the continuum by fitting a linear slope to the central 98~\% channels of each integration, before combining the data together.

%\newpage

\section{Results and discussion}

\subsection{Line detection}

\begin{figure}
 \centering
 \includegraphics[width=8.5cm]{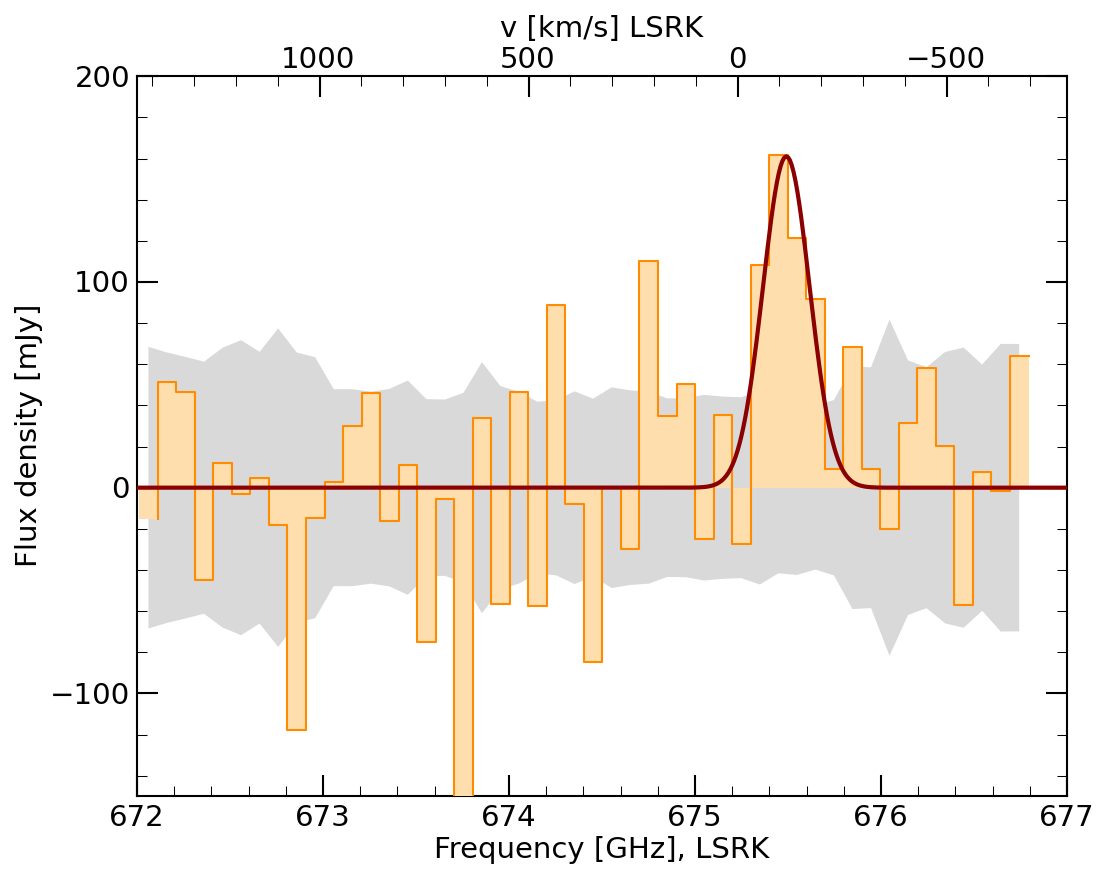}
 \caption{APEX SEPIA~660 spectrum of G09.83808, re-sampled into 100~MHz (45 km s$^{-1}$) bins. The best-fitting Gaussian profile is indicated in red, the grey shading indicates the rms noise. The line is detected at $\sim$4$\sigma$ level over 100~MHz channels, and $\sim$5$\sigma$ over 225~MHz (100~km s$^{-1}$) channels.}
 \label{fig:spectrum}
\end{figure}

We detect the [\ion{O}{1}]$_{63}$ line at 675.45~GHz\footnote{Although APEX observations can not distinguish between the emission from the source and the $z=0.776$ lensing galaxy, our detection does not correspond to any potential emission lines for the foreground lens.} (Figure~\ref{fig:spectrum}), with a peak flux of 2.3$\pm$0.6~mK for 44~km~s$^{-1}$ binning (100~MHz, 3.9$\sigma$ detection), and 2.08$\pm$0.40~mK (5.3$\sigma$) for 100 km~s$^{-1}$ (225~MHz) binning. The line is separately detected at 3.8$\sigma$ (100~MHz bandwidth) in the 2019 October 28 scan. The signal is well-separated from the O$_3$ atmospheric lines at 673.9, 676.1 and 679.3~GHz. We derive the [\ion{O}{1}]$_{63}$ line flux by fitting the combined, continuum-subtracted spectra with a Gaussian profile. 

We processed the data using different channel binning, continuum subtractions and weighting of individual datasets; the line detection is robust against these changes. To account for the atmospheric features, we report the detection with respect to the noise calculated directly from the scatter in the data (gray shading in Figure~\ref{fig:spectrum}, rather than the system temperature from {\sc Gildas}.

Converting the antenna temperature into flux density using the antenna conversion factor of 70~Jy/K, we obtain a line flux of $I_\mathrm{[OI]63}=22\pm5$~Jy km s$^{-1}$, with FWHM=130$\pm$40~km s$^{-1}$. This corresponds to a sky-plane [\ion{O}{1}]$_{63}$ luminosity of $L_\mathrm{[OI]}=(5.4\pm1.2)\times10^{10}$~L$_\odot$. Adjusting for the FIR-based magnification of $\mu_\mathrm{FIR}=9.3\pm1.0$ \citep{Zavala2018}, this translates to a source-plane luminosity of $L_\mathrm{[OI]}=(5.8\pm1.3)\times10^{9}$~L$_\odot$.

We do not measure the rest-frame 63-$\mu$m continuum flux-density, due to the limited total-power stability of the SEPIA~660 receiver.

\subsection{Comparison to [CII] and CO lines}

We now compare our [\ion{O}{1}]$_{63}$ line to the [\ion{C}{2}], CO(6--5) and (5--4) spectra from \citet{Zavala2018}. As all the line observations are unresolved, we assume the same magnification factor as for the FIR continuum. The two-image configuration of G09.83808 limits the effect of differential lensing, as the magnification does not vary dramatically across the source. However, high-resolution studies of $z\geq2$ DSFGs have shown that the [\ion{C}{2}] emission can be substantially more extended than FIR continuum \citep{Gullberg2018, Lamarche2018, Litke2019, Rybak2019, Rybak2019b}, and thus only a fraction of the [\ion{C}{2}] emission might be associated with the [\ion{O}{1}]$_{63}$ and FIR emission.

Compared to the [\ion{C}{2}] luminosity from \citet{Zavala2018}, the [\ion{O}{1}]$_{63}$ line is $\sim$4 times brighter, and 100~times brighter than the CO(6--5)/(5--4) and H$_2$O lines. Therefore, the [\ion{O}{1}]$_{63}$ dominates the gas cooling budget, in agreement with expectations for the dense star-forming ISM in DSFGs \citep{Kaufman1999, Kaufman2006, Narayanan2017}.

\begin{figure}
 \centering
 \includegraphics[width=8.5cm]{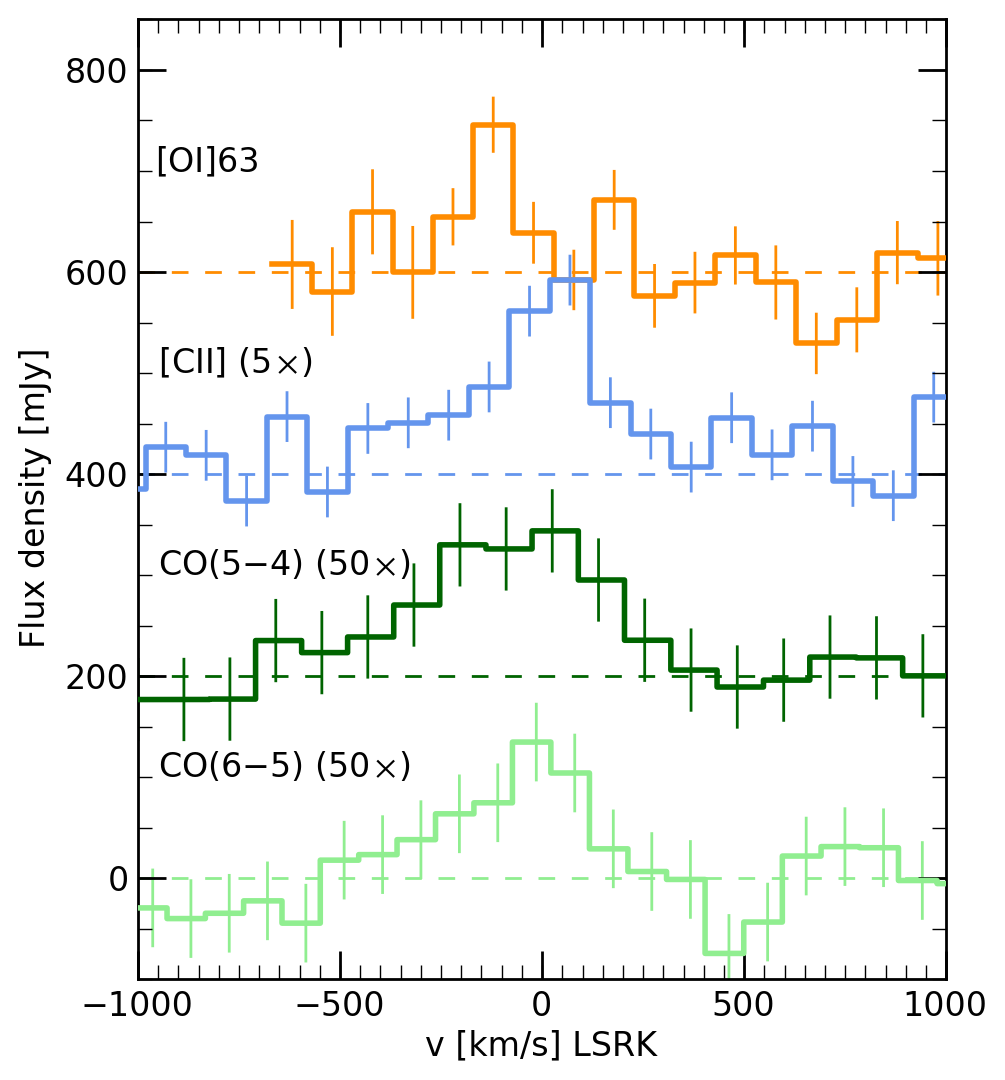}
 \caption{Comparison of the [\ion{O}{1}]$_{63}$ spectrum to the [\ion{C}{2}] and CO(6--5) and (5--4) line profile from \citet{Zavala2018}. The [\ion{O}{1}]$_{63}$] line is noticeably narrower than the [\ion{C}{2}] and CO emission, and tentatively offset from the [\ion{C}{2}] line peak. All spectra have been re-sampled to 100~km s$^{-1}$ bins and are offset by 200~mJy for clarity. The velocities are given in the LSRK frame, using the optical definition.}
 \label{fig:oi_cii_co}
\end{figure}

Figure~\ref{fig:oi_cii_co} compares the [\ion{O}{1}]$_{63}$ line to the [\ion{C}{2}] and CO(6--5)/(5--4) lines from \citet{Zavala2018}. The [\ion{O}{1}]$_{63}$ lines is noticeably narrower than the [\ion{C}{2}] and CO emission (FWHM = 340 - 500 kms$^{-1}$). The centre of the [\ion{O}{1}]$_{63}$ line is consistent with the CO(5--4) and CO(6--5) lines, but offset by $\sim$100~km s$^{-1}$ with respect to the [\ion{C}{2}] line (Figure~\ref{fig:oi_cii_co}). Due to the limited S/N of the data at hand, the variation of [\ion{O}{1}]$_{63}$/[\ion{C}{2}] ratio with velocity remains tentative ($\leq$3$\sigma$ significance). We consider two potential explanations for this discrepancy. First, the [\ion{O}{1}]$_{63}$ emission traces only high-density gas in the central starburst, whereas [\ion{C}{2}] traces the bulk of the gas reservoir, thanks to its much lower critical density ($\sim100$ cm$^{-3}$); the varying [\ion{O}{1}]$_{63}$/[\ion{C}{2}] and [\ion{O}{1}]$_{63}$/CO ratios would then suggest a density gradient across the source. Alternatively, the [\ion{O}{1}]$_{63}$ line might be absorbed in the red channels as seen in some $z\sim0$ ULIRGs (c.f. \citealt{Rosenberg2015, Diaz2017}). A potential [\ion{O}{1}]$_{63}$ self-absorption could be confirmed by comparison with the (much weaker) optically thin [\ion{O}{1}] 145-$\mu$m emission. High-resolution imaging with ALMA and NOEMA will be crucial for disentangling the relative spatial distribution of the [\ion{O}{1}], [\ion{C}{2}], CO and FIR emission.

\subsection{[OI]/FIR and [OI]/[CII] ratios}

\begin{figure}
 \centering
 \includegraphics[width = 8.5cm]{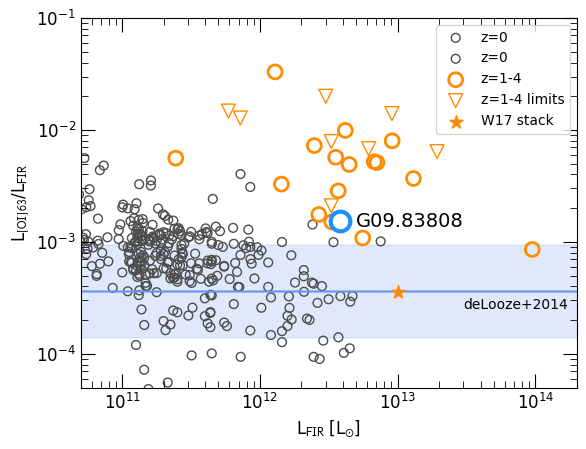}
 \includegraphics[width = 8.5cm]{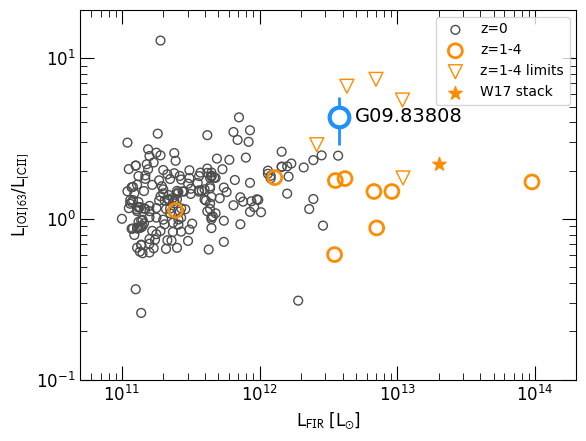}
 \caption{[\ion{O}{1}]/FIR (\textit{upper}) and [\ion{O}{1}]/[\ion{C}{2}] (\textit{lower}) luminosity ratios in G09~83808, compared to other high-redshift detections and upper limits and $z\sim0$ galaxies (GOALS sample from \citealt{Diaz2017} and sources from the \citealt{GraciaCarpio2011} and \citet{Coppin2012} compilation), and the [\ion{O}{1}]$_{63}$-FIR correlation from \citet{deLooze2014}. The line luminosities are given in units of L$_\odot$. FIR luminosities from the literature have been converted to the 8--1000~$\mu$m range. For strongly lensed sources, the luminosities are given as source-plane (de-lensed).}
 \label{fig:oi_fir}
\end{figure}

Figure~\ref{fig:oi_fir} compares the [\ion{O}{1}]$_{63}$/FIR and [\ion{O}{1}]$_{63}$/[\ion{C}{2}] luminosity ratios to literature values for $z\sim0$ galaxies, and $z\geq1$ detections and upper limits. In terms of [\ion{O}{1}]$_{63}$/FIR, G09.83808 is in agreement with $z\sim0$ star-forming galaxies and ULIRGs \citep{Brauher2008, Diaz2017}, contrary to some $z\sim0$ ULIRGs \citep{GraciaCarpio2011, herrera2018b}, the [\ion{O}{1}]$_{63}$ emission in G09.83808 does not show any [\ion{O}{1}]$_{63}$/FIR ''deficit". Compared to the \textit{Herschel} [\ion{O}{1}]$_{63}$ detections, G09.8308 shows a somewhat lower [\ion{O}{1}]$_{63}$/FIR ratio. Rather than indicating that G09.83808 is a special case, this is likely due to a luminosity bias of \textit{Herschel} detections towards [\ion{O}{1}]$_{63}$-luminous sources. For example, all the previous $z\geq1$ [\ion{O}{1}]$_{63}$ detections - apart from the \citet{Wardlow2017} stack - show higher [\ion{O}{1}]$_{63}$/FIR ratio than the star-forming galaxies from the GOALS sample (Figure~\ref{fig:oi_fir}). Comparing the observed [\ion{O}{1}]$_{63}$ luminosity with the FIR-based SFR estimate, G09.83808 falls slightly above the general \citet{deLooze2014} SFR-$L_\mathrm{[OI]63}$ relation, assuming a Salpeter IMF.

The high [\ion{O}{1}]$_{63}$ luminosity also provides an explanation for the observed [\ion{C}{2}] cooling deficit. While the [\ion{C}{2}] line is typically the main coolant of the neutral ISM with [\ion{C}{2}]/FIR ratio of $\sim0.5$\% (comparable to the typical photoelectric heating efficiency), in G09.83808, the observed [\ion{C}{2}]/FIR ratio is $\sim$0.04\%. While the low [\ion{C}{2}]/FIR ratio has been proposed to be a result of lowered photoelectric heating efficiency due to positive grain charging, this does not seem to be the case in G09.83808: the [\ion{O}{1}]$_{63}$ line accounts for $\sim0.16$\% of the total FIR luminosity, and together with the observed [\ion{C}{2}], CO and H$_2$O lines (i.e., notwithstanding any contribution from other cooling lines), this accounts for $\geq0.2$\% of the FIR luminosity, in agreement with standard photoelectric heating models (e.g., \citealt{Bakes1994}). Indeed, G09.83808 has the highest [\ion{O}{1}]/[\ion{C}{2}] ratio among the $z>1$ detections to-date (Figure~\ref{fig:oi_fir}), although consistent with the \citep{Wardlow2017} stack of $z=1-4$ DSFGs within 2$\sigma$. Note that due to the small number of $z>1$ [\ion{O}{1}]$_{63}$ detections, the seven unusually [\ion{C}{2}]-bright sources from the \citet{Brisbin2015} sample ($L_\mathrm{[CII]}/L_\mathrm{FIR}=(0.4-2.0)\times10^{-2}$) bias the high-redshift statistics. As the [\ion{O}{1}]$_{63}$/[\ion{C}{2}] ratio increases with the molecular cloud (surface) density \citep{Narayanan2017}, this suggests a very dense ISM in G09.83808.

\subsection{PDR modelling}
To derive the FUV field strength and density of the neutral ISM from the observed [\ion{O}{1}], [\ion{C}{2}] and FIR luminosities, we use the {\sc PDRToolbox} photon-dominated region models \citep{Kaufman2006, Pound2008}. We adopt the following corrections to the default semi-infinite slab models: (1) as the molecular clouds in DSFGs are likely illuminated both from the front and back, we adjust the {\sc PDRToolbox} predictions for the optical thickness of individuals tracers: while [\ion{C}{2}] and FIR continuum are optically thin and the emission from both the front and back side of the cloud will be detected, the optically thick [\ion{O}{1}]$_{63}$ (and CO) emission will be observed only from the front (c.f., \citealt{Kaufman2006, Diaz2017, Brisbin2015, Rybak2019}; (2) as the [\ion{C}{2}] emission can arise from both neutral and ionized gas, we conservatively adjust the [\ion{C}{2}] luminosity for 20~\% ionized gas contribution (c.f., \citealt{herrera2018b}). We adopt the solar-metallicity {\sc PDRToolbox} model, as FIR indicators point to high ($Z \geq 1~Z_\odot$) metallicity in DSFGs \citep{Wardlow2017}, and as our chosen tracers (FIR, [\ion{O}{1}], [\ion{C}{2}]) are only weakly dependent on $Z$ \citep{Kaufman1999}. 
 
\begin{figure}
 \centering
 \includegraphics[width = 8.5cm]{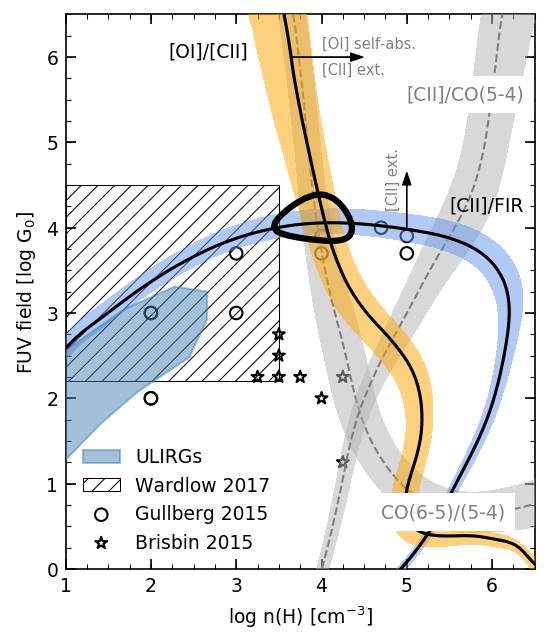}
 \caption{FUV field $G$ and density ($n$) in G09.83808 inferred using the {\sc PDRToolbox} models \citep{Kaufman2006, Pound2008}, compared to other unresolved studies of DSFGs at $z=1-5$ \citep{Brisbin2015, Gullberg2015, Wardlow2017}, and $z\sim0$ ULIRGs from \citet{Diaz2017}. The thick black line indicates the 1$\sigma$ confidence region. The [\ion{C}{2}]/CO(5--4) and CO(6--5)/CO(5--4) line ratios are not used in the PDR modelling. The arrows indicate the direction (not magnitude) of the contours shifting if the [\ion{C}{2}] emission is significantly more extended than FIR continuum, or if [\ion{O}{1}]$_{63}$ is self-absorbed.}
 \label{fig:pdr}
\end{figure}

Figure~\ref{fig:pdr} shows the $G$-$n$ space traced by the observed [\ion{O}{1}]$_{63}$/[\ion{C}{2}] and [\ion{C}{2}]/FIR ratios, in units of L$_\odot$. In terms of an idealized cloud, the [\ion{C}{2}]/FIR is set by $G$ which determines the depth of the C$^+$ layer, while the [\ion{O}{1}]$_{63}$/[\ion{C}{2}] is determined by the gas density.

We obtain a best-fitting model of $G=10^{4.0\pm0.3}~G_0$, $n=10^{4.0\pm0.5}$~cm$^{-3}$. Assuming an optically thin [\ion{O}{1}]$_{63}$ emission shifts the best-fitting $G$ value by $\sim$0.1~dex, while $n$ decreases by $\sim$0.5~dex. Changing the ionized-phase contribution to the [\ion{C}{2}] emission moves the $G$, $n$ values by $\sim$0.1~dex. The derived FUV field and density are comparable to the ISM conditions in $z=1-4$ DSFGs inferred from the fine-structure lines \citep{Wardlow2017} and [\ion{C}{2}] and CO emission \citep{Gullberg2015}, while $\sim$1~dex higher than in $z\sim0$ ULIRGs (\citealt{Diaz2017} and $z\geq1$ source from \citet{Brisbin2015}, inferred from [\ion{C}{2}] and [\ion{O}{1}]$_{63}$). The difference with the \citet{Brisbin2015} sample is mainly due to their high [\ion{C}{2}]/FIR ratios, which determine the $G$ estimates.

Although the CO(5--4) and CO(6--5) lines were excluded from the PDR modelling, the CO(6--5)/(5--4) ratio is consistent with our solution. This is not surprising, as the ratio of the two lines depends mainly on the gas density and is basically unaffected by the CMB \citep{daCunha2013}. On the other hand, the [\ion{C}{2}]/CO(5--4) ratio is offset to much higher densities ($n\simeq10^5$~cm$^{-3}$ for $G=10^4~G_0$). Given the strong dependence of the predicted mid/high-$J$ CO luminosity to the elevated CMB temperature \citep{daCunha2013} which would shift the [\ion{C}{2}]/CO(5--4) isocontour to lower densities, we do not consider this discrepancy to be significant.

If the [\ion{C}{2}] emission is significantly more extended as the FIR continuum, the total [\ion{C}{2}] luminosity associated with the FIR-traced star-forming region will decrease. These would push the PDR model towards higher $G$ and $n$. Similarly, if a significant fraction of the [\ion{O}{1}]$_{63}$ line is self-absorbed, the intrinsic [\ion{O}{1}]$_{63}$ luminosity will increase, moving the best-fitting model to higher densities.

\subsection{Detecting the [OI] 63-um emission from $z \gtrsim6$ DSFGs with ALMA}

What are the prospects of detecting the [\ion{O}{1}]$_{63}$ line from $z\geq6$ non-lensed DSFGs with ALMA Band~9 observations? 

Assuming that the intrinsic (i.e., de-lensed) properties of G09.83808 are representative of the $z\geq6$ DSFG population, i.e. with [\ion{O}{1}]$_{63}$ source-plane luminosity of $5.8\times10^9$~L$_\odot$ over $\sim100$~km s$^{-1}$ linewidth, the [\ion{O}{1}]$_{63}$ emission will be detectable at $\geq$5$\sigma$ level in less than 15~min on-source time. At $z\geq6.8$, the [\ion{O}{1}]$_{63}$ shifts outside the Band~9, and is only redshifted into Band~8 at $z\geq8.5$, when the required on-source time increases into hours. In contrast to G09.83808-like sources, detecting the [\ion{O}{1}]$_{63}$ emission from normal star-forming galaxies such as the population from the \citet{Olsen2017} simulations (SFR = 2-20~$M_\odot$ yr$^{-1}$, $L_\mathrm{[OI]63}=(0.3-2.0)\times10^8~L_\odot$) remains prohibitively expensive.

The modest expense of ALMA time required to detect the [\ion{O}{1}]$_{63}$ emission from G09.83808-like DSFGs will allow an efficient follow-up of $z\gtrsim6$ DSFGs which will be delivered by the on-going and planned mm-wave surveys (e.g., \citealt{Casey2018, Zavala2018b, Gismo2019}). The combination of the [\ion{O}{1}]$_{63}$ and [\ion{C}{2}] emission lines with the FIR continuum will then provide robust measurements of the FUV field and gas density in their star-forming regions.

%\newpage
\section{Conclusions}

We have obtained the first ground-based detection of the [\ion{O}{1}] 63-$\mu$m emission from a $z\geq6$ galaxy, using APEX SEPIA~660 spectroscopy, with only 2:15~h on-source time. This represents the first unambiguous [\ion{O}{1}]$_{63}$ detection beyond redshift 3. In combination with the FIR continuum and [\ion{C}{2}] and CO(6--5)/(5--4) observations from \citet{Zavala2018}, this detection allows us to constrain the physical conditions of the star-forming ISM. Our main findings are:
 \begin{itemize}
 \item The [\ion{O}{1}]$_{63}$ line dominates the neutral gas cooling budget, with a [\ion{O}{1}]/[\ion{C}{2}] ratio of $\sim$4. The shift of the main cooling channel from the [\ion{C}{2}] to the [\ion{O}{1}]$_{63}$ line is in agreement with radiative transfer models of star-forming galaxies (e.g., \citealt{Kaufman1999, Kaufman2006, Narayanan2017, Olsen2017}). The cooling via the [\ion{O}{1}]$_{63}$ line compensates for the pronounced [\ion{C}{2}] deficit in G09.83808; the total [\ion{O}{1}]$_{63}$+[\ion{C}{2}]+CO cooling corresponds to $\geq0.2$\% of the FIR luminosity
 \item The [\ion{O}{1}]$_{63}$ line profile is significantly narrower than the [\ion{C}{2}] and CO(6--5)/(5--4) lines, and blue-shifted by $\sim$100~km s$^{-1}$ with respect to the [\ion{C}{2}] emission. If real, this can be either due to the varying conditions across the source (density in particular), or a self-absorption of the [\ion{O}{1}]$_{63}$ line in the red channels (implying an even higher intrinsic [\ion{O}{1}]$_{63}$ luminosity). Future [\ion{C}{2}] and [\ion{O}{1}]$_{63}$/[\ion{O}{1}]$_{145}$ observations are necessary to distinguish between the two scenarios.
 \item Using the photon-dissociation region models of \citet{Kaufman2006, Pound2008}, we derive a source-averaged FUV field strength $G=10^4~G_0$ and density $n=10^{4.0}$~cm$^{-3}$. These are comparable to source-averaged values for $z=1-4$ DSFG samples, and $\geq$1~dex higher than source-averaged values in $z\sim0$ ULIRGs.
 \item If G09.83808 represents a typical $z\sim6$ DSFG, a 5$\sigma$ detection of the [\ion{O}{1}]$_{63}$ emission from a $z=6$ non-lensed DSFGs will be possible in $\sim15$~min of ALMA Band~9 observations, complementing the currently exploited [\ion{C}{2}] and [\ion{O}{3}] emission.
 \end{itemize}
These results highlight the power of the [\ion{O}{1}] 63-$\mu$m line as a tracer of neutral ISM in DSFGs at the highest redshift. Thanks to its brightness, ground-based studies of the [\ion{O}{1}] 63-$\mu$m line will open a new window into the physics of star-forming neutral ISM in the first billion years of the cosmic history.

\acknowledgements
{The authors thank Kalle Torstensson and Carlos de Breuck for carrying out observations used in this work and their comments on early version of this manuscript. This publication is based on data acquired with the Atacama Pathfinder Experiment (APEX) and the APEX SEPIA receiver, developed by NOVA, the Netherlands Research School for Astronomy. APEX is a collaboration between the Max-Planck-Institut f\"ur Radioastronomie, the European Southern Observatory, and the Onsala Space Observatory. 
MR and JH acknowledge support of the VIDI research programme with project number 639.042.611, which is (partly) financed by the Netherlands Organisation for Scientific Research (NWO). 
CMC thanks the National Science Foundation for support through grants AST-1714528 and AST-1814034, and additionally CMC and JAZ thank the University of Texas at Austin College of Natural Sciences for support. In addition, CMC acknowledges support from the Research Corporation for Science Advancement from a 2019 Cottrell Scholar Award sponsored by IF/THEN, an initiative of Lyda Hill Philanthropies.}

\bibliography{apex_oi}

\begin{thebibliography}{}
\expandafter\ifx\csname natexlab\endcsname\relax\def\natexlab#1{#1}\fi
\providecommand{\url}[1]{\href{#1}{#1}}

\bibitem[{Bakes \& Tielens(1994)}]{Bakes1994}
Bakes, E. L.~O., \& Tielens, A. G. G.~M. 1994, \apj, 427, 822

\bibitem[{{Belitsky} {et~al.}(2018){Belitsky}, {Lapkin}, {Fredrixon},
  {Meledin}, {Sundin}, {Billade}, {Ferm}, {Pavolotsky}, {Rashid}, {Strandberg},
  {Desmaris}, {Ermakov}, {Krause}, {Olberg}, {Aghdam}, {Shafiee}, {Bergman},
  {De Beck}, {Olofsson}, {Conway}, {De Breuck}, {Immer}, {Yagoubov},
  {Montenegro-Montes}, {Torstensson}, {P{\'e}rez-Beaupuits}, {Klein}, {Boland},
  {Baryshev}, {Hesper}, {Barkhof}, {Adema}, {Bekema}, \&
  {Koops}}]{Belitsky2018}
{Belitsky}, V., {Lapkin}, I., {Fredrixon}, M., {et~al.} 2018, \aap, 612, A23

\bibitem[{{Brauher} {et~al.}(2008){Brauher}, {Dale}, \& {Helou}}]{Brauher2008}
{Brauher}, J.~R., {Dale}, D.~A., \& {Helou}, G. 2008, \apjs, 178, 280

\bibitem[{{Brisbin} {et~al.}(2015){Brisbin}, {Ferkinhoff}, {Nikola},
  {Parshley}, {Stacey}, {Spoon}, {Hailey-Dunsheath}, \& {Verma}}]{Brisbin2015}
{Brisbin}, D., {Ferkinhoff}, C., {Nikola}, T., {et~al.} 2015, \apj, 799, 13

\bibitem[{{Carniani} {et~al.}(2017){Carniani}, {Maiolino}, {Pallottini},
  {Vallini}, {Pentericci}, {Ferrara}, {Castellano}, {Vanzella}, {Grazian},
  {Gallerani}, {Santini}, {Wagg}, \& {Fontana}}]{Carniani2017}
{Carniani}, S., {Maiolino}, R., {Pallottini}, A., {et~al.} 2017, \aap, 605, A42

\bibitem[{{Casey} {et~al.}(2014){Casey}, {Narayanan}, \& {Cooray}}]{Casey2014}
{Casey}, C.~M., {Narayanan}, D., \& {Cooray}, A. 2014, \physrep, 541, 45

\bibitem[{{Casey} {et~al.}(2018){Casey}, {Zavala}, {Spilker}, {da Cunha},
  {Hodge}, {Hung}, {Staguhn}, {Finkelstein}, \& {Drew}}]{Casey2018}
{Casey}, C.~M., {Zavala}, J.~A., {Spilker}, J., {et~al.} 2018, \apj, 862, 77

\bibitem[{{Coppin} {et~al.}(2012){Coppin}, {Danielson}, {Geach}, {Hodge},
  {Swinbank}, {Wardlow}, {Bertoldi}, {Biggs}, {Brandt}, {Caselli}, {Chapman},
  {Dannerbauer}, {Dunlop}, {Greve}, {Hamann}, {Ivison}, {Karim}, {Knudsen},
  {Menten}, {Schinnerer}, {Smail}, {Spaans}, {Walter}, {Webb}, \& {van der
  Werf}}]{Coppin2012}
{Coppin}, K.~E.~K., {Danielson}, A.~L.~R., {Geach}, J.~E., {et~al.} 2012,
  \mnras, 427, 520

\bibitem[{{da Cunha} {et~al.}(2013){da Cunha}, {Groves}, {Walter}, {Decarli},
  {Weiss}, {Bertoldi}, {Carilli}, {Daddi}, {Elbaz}, {Ivison}, {Maiolino},
  {Riechers}, {Rix}, {Sargent}, \& {Smail}}]{daCunha2013}
{da Cunha}, E., {Groves}, B., {Walter}, F., {et~al.} 2013, \apj, 766,
  doi:10.1088/0004-637X/766/1/13

\bibitem[{{De Looze} {et~al.}(2014){De Looze}, {Cormier}, {Lebouteiller},
  {Madden}, {Baes}, {Bendo}, {Boquien}, {Boselli}, {Clements}, {Cortese},
  {Cooray}, {Galametz}, {Galliano}, {Graci{\'a}-Carpio}, {Isaak}, {Karczewski},
  {Parkin}, {Pellegrini}, {R{\'e}my-Ruyer}, {Spinoglio}, {Smith}, \&
  {Sturm}}]{deLooze2014}
{De Looze}, I., {Cormier}, D., {Lebouteiller}, V., {et~al.} 2014, \aap, 568,
  A62

\bibitem[{{Decarli} {et~al.}(2017){Decarli}, {Walter}, {Venemans},
  {Ba{\~n}ados}, {Bertoldi}, {Carilli}, {Fan}, {Farina}, {Mazzucchelli},
  {Riechers}, {Rix}, {Strauss}, {Wang}, \& {Yang}}]{Decarli2017}
{Decarli}, R., {Walter}, F., {Venemans}, B.~P., {et~al.} 2017, \nat, 545, 457

\bibitem[{{D{\'{\i}}az-Santos} {et~al.}(2017){D{\'{\i}}az-Santos}, {Armus},
  {Charmandaris}, {Lu}, {Stierwalt}, {Stacey}, {Malhotra}, {van der Werf},
  {Howell}, {Privon}, {Mazzarella}, {Goldsmith}, {Murphy}, {Barcos-Mu{\~n}oz},
  {Linden}, {Inami}, {Larson}, {Evans}, {Appleton}, {Iwasawa}, {Lord},
  {Sanders}, \& {Surace}}]{Diaz2017}
{D{\'{\i}}az-Santos}, T., {Armus}, L., {Charmandaris}, V., {et~al.} 2017, \apj,
  846, 32

\bibitem[{{Graci{\'a}-Carpio} {et~al.}(2011){Graci{\'a}-Carpio}, {Sturm},
  {Hailey-Dunsheath}, {Fischer}, {Contursi}, {Poglitsch}, {Genzel},
  {Gonz{\'a}lez-Alfonso}, {Sternberg}, {Verma}, {Christopher}, {Davies},
  {Feuchtgruber}, {de Jong}, {Lutz}, \& {Tacconi}}]{GraciaCarpio2011}
{Graci{\'a}-Carpio}, J., {Sturm}, E., {Hailey-Dunsheath}, S., {et~al.} 2011,
  \apj, 728, L7

\bibitem[{{Gullberg} {et~al.}(2015){Gullberg}, {De Breuck}, {Vieira},
  {Wei{\ss}}, {Aguirre}, {Aravena}, {B{\'e}thermin}, {Bradford}, {Bothwell},
  {Carlstrom}, {Chapman}, {Fassnacht}, {Gonzalez}, {Greve}, {Hezaveh},
  {Holzapfel}, {Husband}, {Ma}, {Malkan}, {Marrone}, {Menten}, {Murphy},
  {Reichardt}, {Spilker}, {Stark}, {Strandet}, \& {Welikala}}]{Gullberg2015}
{Gullberg}, B., {De Breuck}, C., {Vieira}, J.~D., {et~al.} 2015, \mnras, 449,
  2883

\bibitem[{{Gullberg} {et~al.}(2018){Gullberg}, {Swinbank}, {Smail}, {Biggs},
  {Bertoldi}, {De Breuck}, {Chapman}, {Chen}, {Cooke}, {Coppin}, {Cox},
  {Dannerbauer}, {Dunlop}, {Edge}, {Farrah}, {Geach}, {Greve}, {Hodge}, {Ibar},
  {Ivison}, {Karim}, {Schinnerer}, {Scott}, {Simpson}, {Stach}, {Thomson}, {van
  der Werf}, {Walter}, {Wardlow}, \& {Weiss}}]{Gullberg2018}
{Gullberg}, B., {Swinbank}, A.~M., {Smail}, I., {et~al.} 2018, \apj, 859, 12

\bibitem[{{Harikane} {et~al.}(2019){Harikane}, {Ouchi}, {Inoue}, {Matsuoka},
  {Tamura}, {Bakx}, {Fujimoto}, {Moriwaki}, {Ono}, {Nagao}, {Tadaki}, {Kojima},
  {Shibuya}, {Egami}, {Ferrara}, {Gallerani}, {Hashimoto}, {Kohno}, {Matsuda},
  {Matsuo}, {Pallottini}, {Sugahara}, \& {Vallini}}]{Harikane2019}
{Harikane}, Y., {Ouchi}, M., {Inoue}, A.~K., {et~al.} 2019, arXiv e-prints,
  arXiv:1910.10927

\bibitem[{Hashimoto {et~al.}(2019)Hashimoto, Inoue, Mawatari, Tamura, Matsuo,
  Furusawa, Harikane, Shibuya, Knudsen, Kohno, Ono, Zackrisson, Okamoto,
  Kashikawa, Oesch, Ouchi, Ota, Shimizu, Taniguchi, Umehata, \&
  Watson}]{Hashimoto2019}
Hashimoto, T., Inoue, A.~K., Mawatari, K., {et~al.} 2019, Publications of the
  Astronomical Society of Japan, 71, doi:10.1093/pasj/psz049, 71.
\newblock \url{https://doi.org/10.1093/pasj/psz049}

\bibitem[{{Herrera-Camus} {et~al.}(2018{\natexlab{a}}){Herrera-Camus}, {Sturm},
  {Graci{\'a}-Carpio}, {Lutz}, {Contursi}, {Veilleux}, {Fischer},
  {Gonz{\'a}lez-Alfonso}, {Poglitsch}, {Tacconi}, {Genzel}, {Maiolino},
  {Sternberg}, {Davies}, \& {Verma}}]{Herrera2018}
{Herrera-Camus}, R., {Sturm}, E., {Graci{\'a}-Carpio}, J., {et~al.}
  2018{\natexlab{a}}, ArXiv e-prints, arXiv:1803.04419

\bibitem[{{Herrera-Camus} {et~al.}(2018{\natexlab{b}}){Herrera-Camus}, {Sturm},
  {Graci{\'a}-Carpio}, {Lutz}, {Contursi}, {Veilleux}, {Fischer},
  {Gonz{\'a}lez-Alfonso}, {Poglitsch}, {Tacconi}, {Genzel}, {Maiolino},
  {Sternberg}, {Davies}, \& {Verma}}]{herrera2018b}
---. 2018{\natexlab{b}}, \apj, 861, 95

\bibitem[{{Hesper} {et~al.}(2017){Hesper}, {Khudchenko}, {Baryshev}, {Barkhof},
  \& {Mena}}]{Hesper2017}
{Hesper}, R., {Khudchenko}, A., {Baryshev}, A.~M., {Barkhof}, J., \& {Mena},
  F.~P. 2017, IEEE Transactions on Terahertz Science and Technology, 7, 686

\bibitem[{Hesper {et~al.}(2018)Hesper, Khudchenko, Lindemulder, Bekema,
  de~Haan-Stijkel, Barkhof, Adema, \& Baryshev}]{Hesper2018}
Hesper, R., Khudchenko, A., Lindemulder, M., {et~al.} 2018, in 29th
  International Symposium on Space Terahertz Technology, Pasadena.
\newblock \url{https://www.nrao.edu/meetings/isstt/papers/2018/2018098103.pdf}

\bibitem[{{Hodge} {et~al.}(2013){Hodge}, {Karim}, {Smail}, {Swinbank},
  {Walter}, {Biggs}, {Ivison}, {Weiss}, {Alexander}, {Bertoldi}, {Brandt},
  {Chapman}, {Coppin}, {Cox}, {Danielson}, {Dannerbauer}, {De Breuck},
  {Decarli}, {Edge}, {Greve}, {Knudsen}, {Menten}, {Rix}, {Schinnerer},
  {Simpson}, {Wardlow}, \& {van der Werf}}]{Hodge2013}
{Hodge}, J.~A., {Karim}, A., {Smail}, I., {et~al.} 2013, \apj, 768, 91

\bibitem[{{Inoue} {et~al.}(2016){Inoue}, {Tamura}, {Matsuo}, {Mawatari},
  {Shimizu}, {Shibuya}, {Ota}, {Yoshida}, {Zackrisson}, {Kashikawa}, {Kohno},
  {Umehata}, {Hatsukade}, {Iye}, {Matsuda}, {Okamoto}, \&
  {Yamaguchi}}]{Inoue2016}
{Inoue}, A.~K., {Tamura}, Y., {Matsuo}, H., {et~al.} 2016, Science, 352, 1559

\bibitem[{Ivison {et~al.}(2010)Ivison, Smail, Papadopoulos, Wold, Richard,
  Swinbank, Kneib, \& Owen}]{Ivison2010}
Ivison, R.~J., Smail, I., Papadopoulos, P.~P., {et~al.} 2010, Monthly Notices
  of the Royal Astronomical Society, 404, 198.
\newblock \url{https://doi.org/10.1111/j.1365-2966.2010.16322.x}

\bibitem[{{Katz} {et~al.}(2019){Katz}, {Galligan}, {Kimm}, {Rosdahl},
  {Haehnelt}, {Blaizot}, {Devriendt}, {Slyz}, {Laporte}, \& {Ellis}}]{Katz2019}
{Katz}, H., {Galligan}, T.~P., {Kimm}, T., {et~al.} 2019, \mnras, 487, 5902

\bibitem[{{Kaufman} {et~al.}(2006){Kaufman}, {Wolfire}, \&
  {Hollenbach}}]{Kaufman2006}
{Kaufman}, M.~J., {Wolfire}, M.~G., \& {Hollenbach}, D.~J. 2006, \apj, 644, 283

\bibitem[{{Kaufman} {et~al.}(1999){Kaufman}, {Wolfire}, {Hollenbach}, \&
  {Luhman}}]{Kaufman1999}
{Kaufman}, M.~J., {Wolfire}, M.~G., {Hollenbach}, D.~J., \& {Luhman}, M.~L.
  1999, \apj, 527, 795

\bibitem[{{Kennicutt}(1998)}]{Kennicutt1998}
{Kennicutt}, Robert~C., J. 1998, \araa, 36, 189

\bibitem[{{Lamarche} {et~al.}(2018){Lamarche}, {Verma}, {Vishwas}, {Stacey},
  {Brisbin}, {Ferkinhoff}, {Nikola}, {Higdon}, {Higdon}, \&
  {Tecza}}]{Lamarche2018}
{Lamarche}, C., {Verma}, A., {Vishwas}, A., {et~al.} 2018, \apj, 867, 140

\bibitem[{{Litke} {et~al.}(2019){Litke}, {Marrone}, {Spilker}, {Aravena},
  {B{\'e}thermin}, {Chapman}, {Chen}, {de Breuck}, {Dong}, {Gonzalez}, {Greve},
  {Hayward}, {Hezaveh}, {Jarugula}, {Ma}, {Morningstar}, {Narayanan}, {Phadke},
  {Reuter}, {Vieira}, \& {Weiss}}]{Litke2019}
{Litke}, K.~C., {Marrone}, D.~P., {Spilker}, J.~S., {et~al.} 2019, \apj, 870,
  80

\bibitem[{{Magnelli} {et~al.}(2019){Magnelli}, {Karim}, {Staguhn},
  {Kov{\'a}cs}, {Jim{\'e}nez-Andrade}, {Casey}, {Zavala}, {Schinnerer},
  {Sargent}, {Aravena}, {Bertoldi}, {Capak}, {Riechers}, \&
  {Benford}}]{Gismo2019}
{Magnelli}, B., {Karim}, A., {Staguhn}, J., {et~al.} 2019, \apj, 877, 45

\bibitem[{{Narayanan} \& {Krumholz}(2017)}]{Narayanan2017}
{Narayanan}, D., \& {Krumholz}, M.~R. 2017, \mnras, 467, 50

\bibitem[{{Olsen} {et~al.}(2017){Olsen}, {Greve}, {Narayanan}, {Thompson},
  {Dav{\'e}}, {Niebla Rios}, \& {Stawinski}}]{Olsen2017}
{Olsen}, K., {Greve}, T.~R., {Narayanan}, D., {et~al.} 2017, \apj, 846, 105

\bibitem[{{Planck Collaboration} {et~al.}(2016){Planck Collaboration}, {Ade},
  {Aghanim}, {Arnaud}, {Ashdown}, {Aumont}, {Baccigalupi}, {Banday},
  {Barreiro}, {Bartlett}, \& et~al.}]{Planck2015}
{Planck Collaboration}, {Ade}, P.~A.~R., {Aghanim}, N., {et~al.} 2016, \aap,
  594, A13

\bibitem[{{Popping} {et~al.}(2019){Popping}, {Narayanan}, {Somerville},
  {Faisst}, \& {Krumholz}}]{Popping2019}
{Popping}, G., {Narayanan}, D., {Somerville}, R.~S., {Faisst}, A.~L., \&
  {Krumholz}, M.~R. 2019, \mnras, 482, 4906

\bibitem[{{Pound} \& {Wolfire}(2008)}]{Pound2008}
{Pound}, M.~W., \& {Wolfire}, M.~G. 2008, in Astronomical Society of the
  Pacific Conference Series, Vol. 394, Astronomical Data Analysis Software and
  Systems XVII, ed. R.~W. {Argyle}, P.~S. {Bunclark}, \& J.~R. {Lewis}, 654

\bibitem[{{Riechers} {et~al.}(2013){Riechers}, {Bradford}, {Clements},
  {Dowell}, {P{\'e}rez-Fournon}, {Ivison}, {Bridge}, {Conley}, {Fu}, {Vieira},
  {Wardlow}, {Calanog}, {Cooray}, {Hurley}, {Neri}, {Kamenetzky}, {Aguirre},
  {Altieri}, {Arumugam}, {Benford}, {B{\'e}thermin}, {Bock}, {Burgarella},
  {Cabrera-Lavers}, {Chapman}, {Cox}, {Dunlop}, {Earle}, {Farrah}, {Ferrero},
  {Franceschini}, {Gavazzi}, {Glenn}, {Solares}, {Gurwell}, {Halpern},
  {Hatziminaoglou}, {Hyde}, {Ibar}, {Kov{\'a}cs}, {Krips}, {Lupu}, {Maloney},
  {Martinez-Navajas}, {Matsuhara}, {Murphy}, {Naylor}, {Nguyen}, {Oliver},
  {Omont}, {Page}, {Petitpas}, {Rangwala}, {Roseboom}, {Scott}, {Smith},
  {Staguhn}, {Streblyanska}, {Thomson}, {Valtchanov}, {Viero}, {Wang},
  {Zemcov}, \& {Zmuidzinas}}]{Riechers2013}
{Riechers}, D.~A., {Bradford}, C.~M., {Clements}, D.~L., {et~al.} 2013, \nat,
  496, 329

\bibitem[{{Rosenberg} {et~al.}(2015){Rosenberg}, {van der Werf}, {Aalto},
  {Armus}, {Charmandaris}, {D{\'\i}az-Santos}, {Evans}, {Fischer}, {Gao},
  {Gonz{\'a}lez-Alfonso}, {Greve}, {Harris}, {Henkel}, {Israel}, {Isaak},
  {Kramer}, {Meijerink}, {Naylor}, {Sanders}, {Smith}, {Spaans}, {Spinoglio},
  {Stacey}, {Veenendaal}, {Veilleux}, {Walter}, {Wei{\ss}}, {Wiedner}, {van der
  Wiel}, \& {Xilouris}}]{Rosenberg2015}
{Rosenberg}, M.~J.~F., {van der Werf}, P.~P., {Aalto}, S., {et~al.} 2015, \apj,
  801, 72

\bibitem[{{Rybak} {et~al.}(2019{\natexlab{a}}){Rybak}, {Hodge}, {Vegetti}, {van
  der Werf}, {Andreani}, , {Graziani}, \& {McKean}}]{Rybak2019b}
{Rybak}, M., {Hodge}, J.~A., {Vegetti}, S., {et~al.} 2019{\natexlab{a}},
  arXiv:1912.XXXXX

\bibitem[{{Rybak} {et~al.}(2019{\natexlab{b}}){Rybak}, {Calistro Rivera},
  {Hodge}, {Smail}, {Walter}, {van der Werf}, {da Cunha}, {Chen},
  {Dannerbauer}, {Ivison}, {Karim}, {Simpson}, {Swinbank}, \&
  {Wardlow}}]{Rybak2019}
{Rybak}, M., {Calistro Rivera}, G., {Hodge}, J.~A., {et~al.}
  2019{\natexlab{b}}, \apj, 876, 112

\bibitem[{{Smit} {et~al.}(2018){Smit}, {Bouwens}, {Carniani}, {Oesch},
  {Labb{\'e}}, {Illingworth}, {van der Werf}, {Bradley}, {Gonzalez}, {Hodge},
  {Holwerda}, {Maiolino}, \& {Zheng}}]{Smit2018}
{Smit}, R., {Bouwens}, R.~J., {Carniani}, S., {et~al.} 2018, \nat, 553, 178

\bibitem[{{Stacey} {et~al.}(2010){Stacey}, {Hailey-Dunsheath}, {Ferkinhoff},
  {Nikola}, {Parshley}, {Benford}, {Staguhn}, \& {Fiolet}}]{Stacey2010}
{Stacey}, G.~J., {Hailey-Dunsheath}, S., {Ferkinhoff}, C., {et~al.} 2010, \apj,
  724, 957

\bibitem[{{Strandet} {et~al.}(2017){Strandet}, {Weiss}, {De Breuck}, {Marrone},
  {Vieira}, {Aravena}, {Ashby}, {B{\'e}thermin}, {Bothwell}, {Bradford},
  {Carlstrom}, {Chapman}, {Cunningham}, {Chen}, {Fassnacht}, {Gonzalez},
  {Greve}, {Gullberg}, {Hayward}, {Hezaveh}, {Litke}, {Ma}, {Malkan}, {Menten},
  {Miller}, {Murphy}, {Narayanan}, {Phadke}, {Rotermund}, {Spilker}, \&
  {Sreevani}}]{Strandet2017}
{Strandet}, M.~L., {Weiss}, A., {De Breuck}, C., {et~al.} 2017, \apjl, 842, L15

\bibitem[{{Sturm} {et~al.}(2010){Sturm}, {Verma}, {Graci{\'a}-Carpio},
  {Hailey-Dunsheath}, {Contursi}, {Fischer}, {Gonz{\'a}lez-Alfonso},
  {Poglitsch}, {Sternberg}, {Genzel}, {Lutz}, {Tacconi}, {Christopher}, \& {de
  Jong}}]{Sturm2010}
{Sturm}, E., {Verma}, A., {Graci{\'a}-Carpio}, J., {et~al.} 2010, \aap, 518,
  L36

\bibitem[{{Wardlow} {et~al.}(2017){Wardlow}, {Cooray}, {Osage}, {Bourne},
  {Clements}, {Dannerbauer}, {Dunne}, {Dye}, {Eales}, {Farrah}, {Furlanetto},
  {Ibar}, {Ivison}, {Maddox}, {Micha{\l}owski}, {Riechers}, {Rigopoulou},
  {Scott}, {Smith}, {Wang}, {van der Werf}, {Valiante}, {Valtchanov}, \&
  {Verma}}]{Wardlow2017}
{Wardlow}, J.~L., {Cooray}, A., {Osage}, W., {et~al.} 2017, \apj, 837,
  doi:10.3847/1538-4357/837/1/12

\bibitem[{{Wright}(2006)}]{Wright2006}
{Wright}, E.~L. 2006, \pasp, 118, 1711

\bibitem[{{Yang} {et~al.}(2019){Yang}, {Gavazzi}, {Beelen}, {Cox}, {Omont},
  {Lehnert}, {Gao}, {Ivison}, {Swinbank}, {Barcos-Mu{\~n}oz}, {Neri}, {Cooray},
  {Dye}, {Eales}, {Fu}, {Gonz{\'a}lez-Alfonso}, {Ibar}, {Micha{\l}owski},
  {Nayyeri}, {Negrello}, {Nightingale}, {P{\'e}rez-Fournon}, {Riechers},
  {Smail}, \& {van der Werf}}]{Yang2019}
{Yang}, C., {Gavazzi}, R., {Beelen}, A., {et~al.} 2019, \aap, 624, A138

\bibitem[{{Zavala} {et~al.}(2018{\natexlab{a}}){Zavala}, {Casey}, {da Cunha},
  {Spilker}, {Staguhn}, {Hodge}, \& {Drew}}]{Zavala2018b}
{Zavala}, J.~A., {Casey}, C.~M., {da Cunha}, E., {et~al.} 2018{\natexlab{a}},
  \apj, 869, 71

\bibitem[{{Zavala} {et~al.}(2018{\natexlab{b}}){Zavala}, {Monta{\~n}a},
  {Hughes}, {Yun}, {Ivison}, {Valiante}, {Wilner}, {Spilker}, {Aretxaga},
  {Eales}, {Avila-Reese}, {Ch{\'a}vez}, {Cooray}, {Dannerbauer}, {Dunlop},
  {Dunne}, {G{\'o}mez-Ruiz}, {Micha{\l}owski}, {Narayanan}, {Nayyeri}, {Oteo},
  {Rosa Gonz{\'a}lez}, {S{\'a}nchez-Arg{\"u}elles}, {Schloerb}, {Serjeant},
  {Smith}, {Terlevich}, {Vega}, {Villalba}, {van der Werf}, {Wilson}, \&
  {Zeballos}}]{Zavala2018}
{Zavala}, J.~A., {Monta{\~n}a}, A., {Hughes}, D.~H., {et~al.}
  2018{\natexlab{b}}, Nature Astronomy, 2, 56

\bibitem[{{Zhang} {et~al.}(2018){Zhang}, {Ivison}, {George}, {Zhao}, {Dunne},
  {Herrera-Camus}, {Lewis}, {Liu}, {Naylor}, {Oteo}, {Riechers}, {Smail},
  {Yang}, {Eales}, {Hopwood}, {Maddox}, {Omont}, \& {van der
  Werf}}]{Zhang2018b}
{Zhang}, Z.-Y., {Ivison}, R.~J., {George}, R.~D., {et~al.} 2018, \mnras, 481,
  59

\end{thebibliography}
\end{document}